\documentclass[a4paper]{jpconf}
\pdfoutput=1 
\usepackage[dvipdfmx]{graphicx}
\usepackage{booktabs}
\usepackage{multirow}
\usepackage{amsmath}
\usepackage{amsfonts}
\usepackage[para,online,flushleft]{threeparttable}
\usepackage{amssymb}
\usepackage{bm}
\usepackage{tikz}
\usepackage{layouts}
\usepackage{caption}
\usepackage{subcaption}
\usepackage[style=ext-numeric,doi=false,isbn=false,url=false,eprint=false,sorting=none]{biblatex} 
\DeclareFieldFormat[article,periodical]{volume}{\mkbibbold{#1}}

\begin{document}
\title{Immersed-Boundary Fluid-Structure Interaction of Membranes and Shells}

\author{Marin Lauber$^{1,2}$, Gabriel D. Weymouth$^{2}$, Georges Limbert$^{1,3}$}
\address{$^1$Univeristy of Southampton, Southampton, UK\\$^2$Delft University of Technology, Delft, Netherlands\\$^3$University of Cape Town, Cape Town, South Africa}
\ead{m.lauber@tudelft.nl}

\begin{abstract}
This paper presents a general and robust method for the fluid-structure interaction of membranes and shells undergoing large displacement and large added-mass effects by coupling an immersed-boundary method with a shell finite-element model. The immersed boundary method can accurately simulate the fluid velocity and pressure induced by dynamic bodies undergoing large displacements using a computationally efficient pressure projection finite volume solver. The structural solver can be applied to bending and membrane-related problems, making our partitioned solver very general. We use a strongly-coupled algorithm that avoids the expensive computation of the inverse Jacobian within the root-finding iterations by constructing it from input-output pairs of the coupling variables from the previous time steps. Using two examples with large deformations and added mass contributions, we demonstrate that the resulting quasi-Newton scheme is stable, accurate, and computationally efficient.
\end{abstract}

\section{Introduction}\label{sec:intro}
% textwidth in cm: \printinunitsof{cm}\prntlen{\textwidth}
Fluid-structure interaction of thin, flexible structures is omnipresent in nature and engineering, from leaf fluttering to parachute inflation. The low bending rigidity and the small thickness of the structure result in substantial deformations and significant added-mass effects, leading to significant challenges in their simulation. 

First, the large deformations typical to immersed membranes and shells are best treated with Cartesian grid methods. These methods allow for arbitrary displacements and deformations and are computationally attractive. However, the boundary conditions must be enforced correctly for accurate fluid loading, and this can be difficult when immersing thin structures on a Cartesian grid \cite{Lauber2022}. 

Second, coupling the fluid and shell models requires finding an equilibrium between the instantaneous forces and displacements at each point along the interface. Monolithic approaches solve the two systems simultaneously, but generalising the coupling then becomes problematic \cite{Causin2005}. Partitioned solvers are more general, but the large deformations and significant added-mass for thin structures results in a mathematically stiff dynamic system, making it difficult for partitioned solvers to find the instantaneous equilibrium condition. Jacobi or Gauss-Seidel iterations between a fluid and structural solver within a time step converge very slowly, especially for incompressible flows \cite{Degroote2013}. Under-relaxation of the fixed-point iterations can mitigate instabilities and accelerate convergence when added-mass effects are small. For stiffer problems, Aitken's dynamic relaxation, where the relaxation parameter is adjusted using the coupling residuals \cite{kuttler2008}, can speed up the solution while maintaining the stability of the scheme. 

Due to the fast convergence rates, Newton-Raphson methods  \cite{Gerbeau2003, Fernandez2004} are attractive for solving strongly-coupled problems where small relaxation parameters are required with Jacobi or Gauss-Seidel methods. Newton-Raphson methods reformulates the fixed-point problem into a root-finding problem for the interface displacements; standard iterative methods (GMRES, for example) can be used to approximate the linear problem in the Newton step. Within those iterations, each matrix-vector product can be evaluated using finite differences. However, this requires evaluating the fluid and structural solver in each Krylov iteration of each Newton iteration, which is extremely expensive. Reusing the Krylov vector within a time step to improve computational efficiency is possible \cite{Michler2005}. Alternatively, the Jacobian can be estimated purely from reduced-order models \cite{Vierendeels2007}, but these models are often problem-specific and thus limit the generality of the solver.

In this article, we present a novel immersed boundary fluid-structure interaction solver to deal with problems induced by $i)$ thin, flexible structure, $ii)$ large deformation and $iii)$ strong added-mass effects. This article is organized as follows; \sref{sec:immersed} briefly introduces our immersed boundary solver for thin dynamic structures. In \sref{sec:shell}, we describe the shell model and the discretisation method used to solve the structural equations. The coupling approach and the quasi-Newton scheme used to solve the fixed-point problem are presented in \sref{sec:IQNILS}. Finally, we provide three examples of applications of our coupled solver to fluid-structure interaction systems of various complexity in \sref{sec:VANDV}.

\section{Strongly-Coupled Partitioned Simulation}

To solve problems under large deformations and strong added-mass effects, we couple our immersed boundary method to a three-dimensional (3D) shell model. Using the partitioned approach allows us to generalise the coupling to various types of structures but requires strongly coupled partitioned schemes.

\subsection{Immersed Boundary Method}\label{sec:immersed}

Immersed boundary methods are extremely well suited to model large deformation, as opposed to body-fitted methods that require a constant deformation of the domain. However, we have shown in \cite{Lauber2022} that most immersed boundary methods result in fluid leakage across the interface for vanishing-thickness bodies. This leakage results from an inaccurate (or inexistent) pressure boundary condition on the Poisson equation during the projection step. In \cite{Lauber2022}, we introduced an extension to the Boundary Data Immersion Method, which we called BDIM-$\sigma$, that allows for treating thin, dynamic, immersed bodies. The method relies on a convolution of the solid and fluid equations onto a single formulation, valid over the whole domain. This corrects the fluid leakage by explicitly imposing a consistent Neumann boundary condition on the pressure during the projection step. The method adjusts the kernel width and the signed distance function to the immersed body to ensure that the immersed boundary is discredited correctly and that the boundary condition is accurately imposed on the immersed body. Because the method relies on a variable coefficient Poisson equation to determine the pressure, standard algebra methods can be used, making the method extremely effective. Details of the flow solver and validation of bluff and thin bodies can be found in \cite{Maertens2015, Lauber2022}.

\subsection{Shell Structural Solver}\label{sec:shell}

To model the response of the thin structure, we use a 3D shell formulation, see \cite{chapelle2010finite} for details. Compared to standard \emph{Kirchhoff-Love} of \emph{Reissner-Mindlin} shell formulations, the 3D formulation drops the plane stress assumption and accounts for trough thickness variations of the kinematic and kinetic response. This formulation is identical to a standard 3D continuum formulation, which is particularly convenient for non-linear material analysis, as general 3D material law can be directly injected into this formulation. This formulation is implemented in the open-source finite element solver \emph{CalculiX}; details can be found in \cite{Dhondt2004}.

\subsection{Interface Quasi-Newton Method}\label{sec:IQNILS}

For strong coupled physical systems, the fast convergence rate of Newton-Raphson methods is attractive for partitioned simulation. However, a finite-difference approximation of the Jacobian is sub-optimal due to the multiple evaluations of the residual operator. Quasi-Newton methods have recently received attention due to the theoretical fast convergence rates and the reduced cost associated with evaluating the interface Jacobian \cite{Bogaers2014, Degroote2009, Degroote2013}. These make them strong candidates for partitioned approaches, where only the coupling data is used to couple the systems through the fixed-point problem
\begin{equation}\label{resid}
    \mathcal{S}\circ\mathcal{F} ({\bf d}_\Gamma) - {\bf d}_\Gamma \equiv \mathcal{R}({\bf d}_\Gamma)= 0,
\end{equation}
where ${\bf d}_\Gamma\in\mathbb{R}^m$ is the displacement vector of the $m$ degrees of freedom on the interface and $\mathcal{S}$, $\mathcal{F}$ are the structural and fluid operators. The dot implies function composition. $\mathcal{R}$ is the residual operator, and we seek the displacement, which is a root of this operator at each time step.

We solve this root finding problem using the Interface Quasi-Newton Least-Square (IQN-ILS) approach, detailed in \cite{Degroote2009}. Briefly, the Newton method for equation~\eqref{resid} estimates the displacement at step $k+1$ as
\begin{equation}
    {\bf d}_\Gamma^{k+1} = {\bf d}_\Gamma^{k}+ \Delta {\bf d}^k_\Gamma = {\bf d}_\Gamma^{k}-J^k\backslash \mathcal{R}^k
\end{equation}
where $J^k=\frac{\partial \mathcal{R}}{\partial {\bf d}_\Gamma}|_k$ is the $m\times m$ Jacobian of the residual operator at step $k$ and $\backslash$ indicates a left matrix division. The construction of this operator and matrix division would be extremely expensive, and IQN-ILS uses a sequence of $n\ll m$ examples of previous displacements updates and residuals
\begin{equation}
    \bm{W}^k = \left[\Delta{\bf d}_{\Gamma}^{k},\,\,\Delta{\bf d}_{\Gamma}^{k-1},\ldots,\,\,\Delta{\bf d}_{\Gamma}^{k-n+1}\right], \quad
    \bm{V}^k = \left[\mathcal{R}^k, \mathcal{R}^{k-1},\ldots,\mathcal{R}^{k-n+1}\right]
\end{equation}
to estimate the action of the Jacobian as
\begin{equation}
    \Delta {\bf d}^k_\Gamma = -J^k\backslash\mathcal{R}^k = \bm{W}^k\left({\bm{V}^k}^\top\bm{V}^k\right)\backslash{\bm{V}^k}^\top\mathcal{R}^k.
\end{equation}
This represents a least-squares approximation to the Newton solution, which has several advantages. First, the matrices containing the displacements and corresponding residuals are available from previous iterations. Second, the new (linear) problem is for an $n\times n$ matrix and not the full $m\times m$ matrix. Third, methods such as QR decomposition can be used to efficiently increment the RHS of (\theequation) at each iteration, as $\bm{V}^k$ and $\bm{W}^k$ are incremented at each iteration.

Our coupled solver implements this update via the \emph{preCICE} coupling library \cite{preCICEv2}. To the best of our knowledge, this is the first coupling of an immersed boundary method specifically derived for thin structures to a 3D shell model via a quasi-Newton method. BDIM-$\sigma$ allows us to treat of zero-thickness deforming bodies accurately, and the quasi-Newton coupling mitigates the added-mass instabilities inherent to fluid-structure interaction in incompressible flows. Finally, the 3D shell model is agnostic to the shell thickness.

\section{Numerial Examples}\label{sec:VANDV}

We now present two challenging numerical examples of fluid-structure interaction systems undergoing large displacement and under large added-mass effects. We first study the behaviour of a membrane wing at small angles of attack with mild wake unsteadiness. We then test our solver against the challenging cases of vanishing mass (or vanishing mass ratio $M_\rho=\rho_s h/(\rho_f L)\rightarrow{}0$) and vanishing stiffness (vanishing Cauchy number $Ca=Eh^3/(12(1-\nu_s)^2\rho_f U^2 L^3)\rightarrow{}0$) with a flexible flag in the inverted configuration i.e. fixed at the trailing edge, undergoing different flapping regimes. Finally, we compare the quasi-Newton algorithm with standard Gauss-Seidel relaxation.

\subsection{Membrane-Wing}\label{sec:membrane_wing}

The flow around a very thin flexible membrane airfoil has been investigated experimentally \cite{Rojratsirikul2009} and numerically \cite{Gordnier2009}. We replicate the system described in \cite{Gordnier2009} of a flexible membrane of length $L$ and thickness $h$ immersed in a viscous fluid flowing at a velocity $U$. Because their influence on the time-averaged fluid-structure interaction response of the system is minimal (even at a high Reynolds number ($Re=10000$) \cite{Serrano2016}), we omit the two rigid streamlined support at the trailing and leading edge. The relative membrane elasticity is set to $Ca =7.4\times10{-6}$, and the mass ratio to $M_\rho=0.589$. We adopt the Reynolds number $Re=UL/\nu=2500$ of \cite{Gordnier2009} for our two-dimensional (2D) simulation instead of the actual Reynolds number of the experiments, which would require 3D simulations.

The fluid domain consists of a uniform region around the membrane (located at the origin) of dimensions $[-0.5L, 1.5L]\times[-0.5L,0.5L]$. Grid stretching is then used to fill the remaining space until the total domain reaches a size of $[-8L, 16L]\times[-8L,8L]$. The uniform part of mesh as dimensions $\Delta x = \Delta y = L/128$ resulting in a non-dimensional wall distance $y^+=1.25$, assuming a $1/7$ power-law for the skin friction coefficient and using the local grid size $\Delta x=\Delta y$ as the normal wall distance. The membrane is discretised with $128$ linear strain triangular shell elements along the length of the membrane, resulting in a structural grid spacing $\Delta s= \Delta x$. The shell thickness is set to replicate the aspect ratio $L/h=750$ of the physical model. A linear isotropic elastic model is used for the structural response of the membrane with a Poisson's ratio $\nu=0.0$ to compare the results with the high-resolution numerical simulations of \cite{Gordnier2009}. We use a constant time step of $\Delta t U/ \Delta x =0.2$ with initial relaxation parameter $\omega=0.05$ for the Jacobi steps. We let the solution evolve until a quasi-steady state is reached, which takes around 10 convective times. After that, the simulation is carried out another 100 convective times. Upon reaching the quasi-steady state, the solver converges to residuals below $\varepsilon_{rel}=10 ^{-4}$ within each time-steps with an average of 3.05 iteration per time step.

\subsubsection{Convergence} 

\begin{figure}
    \centering
    \def\svgwidth{\textwidth}
    \graphicspath{{Figures/}}
    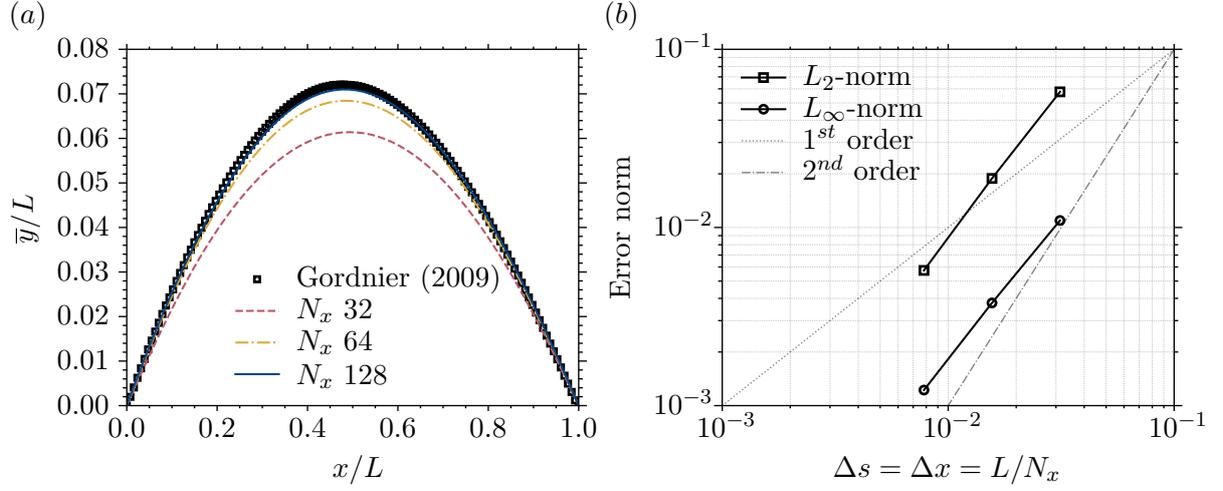
    \caption{$(a)$ Time-averaged deformation of a membrane pinned at the leading and trailing edge in a uniform flow at $Re=2500$, $Ca=7.4\times10{-6}$ and $M_\rho=0.589$ at an angle of attack of 4.0$^\circ$ for different mesh sizes and $(b)$ error norm compared to the reference data from \cite{Gordnier2009}.}
    \label{fig1}
\end{figure}

We use the data from \cite{Gordnier2009} as a reference solution and compute the $L_2$- and $L_\infty$-norm of the difference between the reference data and the deformation obtained at a particular resolution. We vary the fluid and structural resolutions simultaneously within $L/\Delta x = L/\Delta s \in[32,64,128]$, such that $\Delta s/\Delta x$ remains constant, and compute the solution for 150 convective times and then average the last 100 convective times, see \fref{fig1}$a$. We observe monotonic convergence to the reference data for the two error norms, with respective apparent order of convergence of $p=1.56$ and $p=1.49$ for the $L_2$- and $L_\infty$-norm, respectively, see \fref{fig1}$b$.

The nearly quadratic convergence of the scheme for this quasi-steady problem is due to both the fluid and structural solver displaying second-order convergence in space. However, as demonstrated in \cite{Lauber2022}, both geometrical and numerical convergence occur simultaneously in this case with our immersed boundary method.

\subsubsection{Results}

\begin{figure}
    \centering
    \def\svgwidth{\textwidth}
    \graphicspath{{Figures/}}
    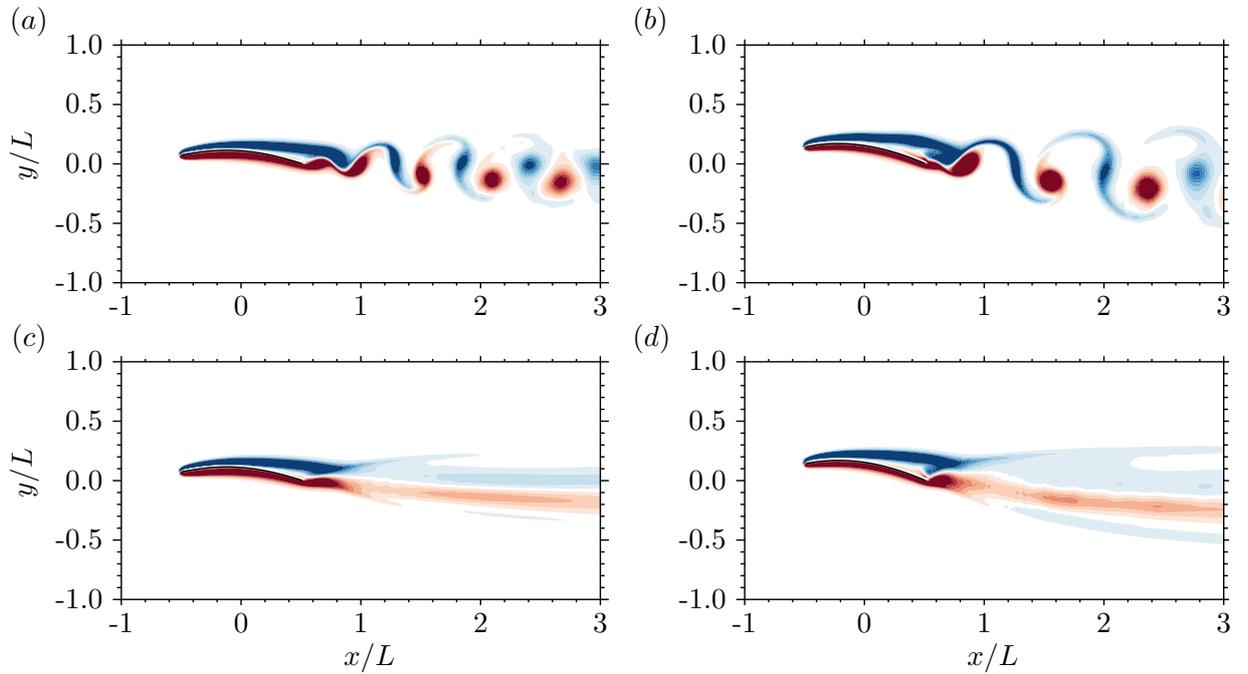
    \caption{Instantaneous (a,b) and time-averaged (c,d) vorticity contours of a uniform flow past a flexible membrane pinned at the leading and trailing edge at $Re=2500$, $Ca=7.4\times10{-6}$ and $M_\rho=0.589$ at an angle of attack of $(a,c)$ 4.0$^\circ$ and $(b,d)$ 8.0$^\circ$. The large transient period at the start (first 10 convective times) has been removed before performing the averaging. Vorticity contours are uniformly spaced in $\omega U / L\in [-8,8]$ in 18 increments.}
    \label{fig2}
\end{figure}

\begin{figure}
    \centering
    \def\svgwidth{\textwidth}
    \graphicspath{{Figures/}}
    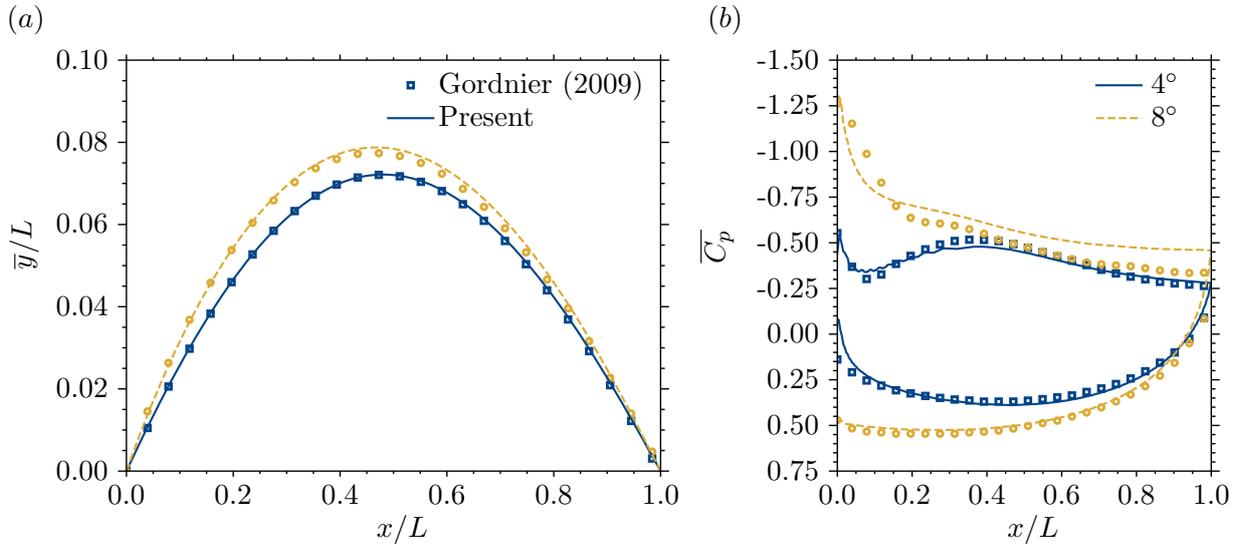
    \caption{Membrane pinned at the leading and trailing edge at an angle of attack $\alpha=4^\circ,8^\circ$ in a uniform flow at $Re=2500$, $Ca=7.4\times10{-6}$ and $M_\rho=0.589$; $(a)$ time-averaged deformation and $(b)$ time-averaged pressure distribution on the upper and lower surface of the membrane. Continuous and dashed lines represent (present) simulation results, square and circular markers represent reference data \cite{Gordnier2009}. Blue and yellow lines/markers represent results at angle-of-attacks $4^\circ$ and $8^\circ$, respectively.}
    \label{fig3}
\end{figure}

Figure~\ref{fig2} show the instantaneous and time-averaged flow around the membrane wing at two angles of attack. Both angles of attack generate a periodic vortex shedding in the wake, generating small amplitude vibration in the membrane. The frequency of oscillations is locked in with the wake frequency. Time-average deformation for the wing for the two angles of attack is presented in \fref{fig3}$a$. 

To demonstrate the coupled solver's ability to resolve pressure forces accurately, we present the upper and lower surface time-averaged pressure coefficient distribution on the membrane wing at two different angles of attack, four and eight degrees, see \fref{fig3}$b$. The pressure on the underside of the membrane agrees well with the reference data for both angles of attack, as the mounts have negligible influence on the pressure field on the underside of the membrane. However, the pressure distribution on the airfoil's upper surface differs from the reference data. We attribute these differences to the influence of the mounts of the membrane omitted in our simulation, which significantly affects the wake and the pressure field on the membrane, especially near the leading edge of the membrane, where a separation bubble is present. While the pressure field shows significant differences, the deformed shape of the membrane is similar regardless of the mount shape \cite{Serrano2016}.

\subsection{Inverted Flag}\label{sec:inverted_flag}

\begin{figure}
    \centering
    \def\svgwidth{\textwidth}
    \graphicspath{{Figures/}}
    %% Creator: Inkscape 1.2.2 (1:1.2.2+202212051552+b0a8486541), www.inkscape.org
%% Accompanies image file 'Figure_4a.pdf' (pdf, eps, ps)
\begingroup%
  \makeatletter%
  \providecommand\color[2][]{%
    \errmessage{(Inkscape) Color is used for the text in Inkscape, but the package 'color.sty' is not loaded}%
    \renewcommand\color[2][]{}%
  }%
  \providecommand\transparent[1]{%
    \errmessage{(Inkscape) Transparency is used (non-zero) for the text in Inkscape, but the package 'transparent.sty' is not loaded}%
    \renewcommand\transparent[1]{}%
  }%
  \providecommand\rotatebox[2]{#2}%
  \newcommand*\fsize{\dimexpr\f@size pt\relax}%
  \newcommand*\lineheight[1]{\fontsize{\fsize}{#1\fsize}\selectfont}%
  \ifx\svgwidth\undefined%
    \setlength{\unitlength}{408.00489699bp}%
    \ifx\svgscale\undefined%
      \relax%
    \else%
      \setlength{\unitlength}{\unitlength * \real{\svgscale}}%
    \fi%
  \else%
    \setlength{\unitlength}{\svgwidth}%
  \fi%
  \global\let\svgwidth\undefined%
  \global\let\svgscale\undefined%
  \makeatother%
  \begin{picture}(1,0.28725937)%
    \lineheight{1}%
    \setlength\tabcolsep{0pt}%
    \put(0,0){\includegraphics[width=\unitlength,page=1]{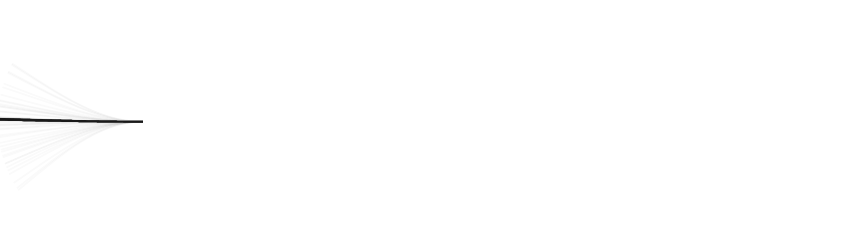}}%
    \put(0.03621474,0.26606462){\makebox(0,0)[rt]{\lineheight{0}\smash{\begin{tabular}[t]{r}III\end{tabular}}}}%
    \put(0,0){\includegraphics[width=\unitlength,page=2]{Figure_4a.pdf}}%
    \put(0.31194672,0.26606462){\makebox(0,0)[rt]{\lineheight{0}\smash{\begin{tabular}[t]{r}IV\end{tabular}}}}%
    \put(0,0){\includegraphics[width=\unitlength,page=3]{Figure_4a.pdf}}%
    \put(0.58767874,0.26606462){\makebox(0,0)[rt]{\lineheight{0}\smash{\begin{tabular}[t]{r}V\end{tabular}}}}%
    \put(0,0){\includegraphics[width=\unitlength,page=4]{Figure_4a.pdf}}%
    \put(0.86341072,0.26606462){\makebox(0,0)[rt]{\lineheight{0}\smash{\begin{tabular}[t]{r}VI\end{tabular}}}}%
  \end{picture}%
\endgroup%

    \caption{Schematic of the different flapping modes of an inverted flag of mass $M_\rho=0.5$ and $Re=200$ for various stiffness $Ca$; (III) small amplitude limit-cycle flapping, (IV) large amplitude limit-cycle flapping, (V) chaotic flapping and (VI) deflected mode flapping. The undeformed equilibrium mode (I) and the deformed equilibrium (II) are not shown.}
    \label{fig4a}
\end{figure}

In this section, we present the problem of a flexible flag pinned at the trailing edge and free at the leading edge. This inverted flag problem has been studied analytically \cite{Rojratsirikul2009}, experimentally \cite{Sader2016} and numerically \cite{Goza2018}, and its behaviour is well documented. In particular, the different flapping regimes and their transition are well established; see \fref{fig4a}. The set-up consists of a flexible flag of length $L$, thickness $h$ and $L/h=50$, immersed in a viscous fluid. We focus on flags with very low mass ratios  ($M_\rho=0.5$) and $Re=200$ for Cauchy numbers spanning the whole range of flapping regimes.

We perform 2D simulations of this system with a fluid domain identical to the one used for the membrane wing. The uniform part of the mesh has a mesh size $\Delta x= \Delta y = L/128$. This gives a local wall non-dimensional distance of $y^+=0.12$. The flag is modelled as a shell of thickness $h$, clamped at the trailing edge and uses a discretisation identical to the membrane wing case presented above, $\Delta s=L/128$. The time-step of the simulations is held constant at $\Delta t U/ \Delta x = 0.2$, where $U$ is the free-stream velocity. An external force is applied to the flag during the first convective time to trigger flapping. After that, the system evolves for 400 convective times before statistics are gathered. The solution to the coupled problem is obtained with an average of 3.56 IQL-ILS iterations per time step.

\begin{figure}
    \centering
    \def\svgwidth{\textwidth}
    \graphicspath{{Figures/}}
    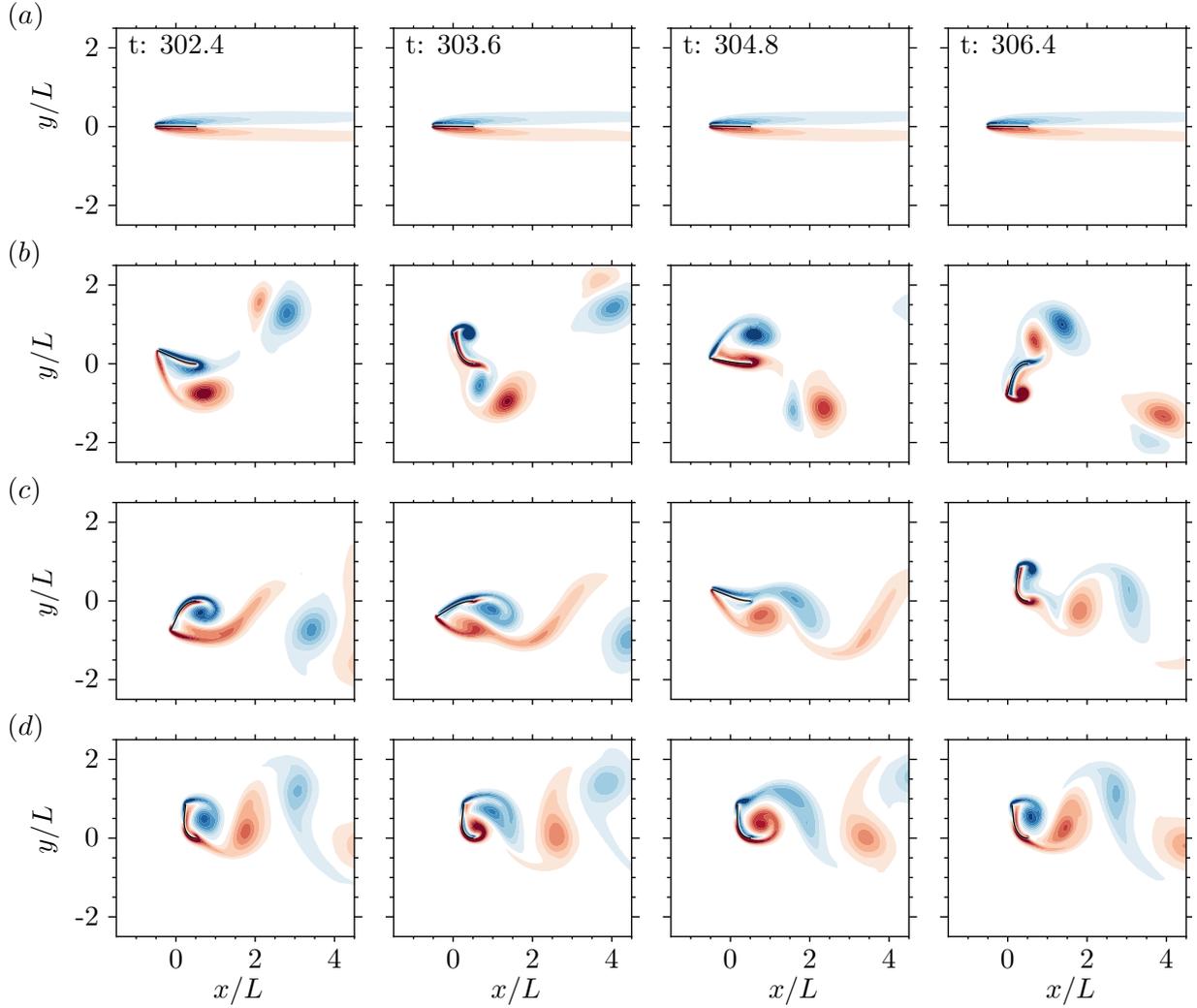
    \caption{Inverted flag of mass $M_\rho=0.5$ and $Re=200$ for various stiffness $Ca$; vorticity contours of the flow around the inverted flag at different snapshots in time, $(b)$ $Ca=0.407$, $(c)$ $Ca=0.350$,  $(d)$ $Ca=0.164$ and $(e)$ $Ca=0.107$. The snapshots are taken at equal times for each Cauchy number.  Vorticity contours are equally spaced between $\omega U/L\in[-10, 10]$ in 18 increments.}
    \label{fig4}
\end{figure}

\subsubsection{Results}

Instantaneous contours of vorticity at selected Cauchy number (same as the one displayed in \fref{fig4a}) are presented in \fref{fig4}. The flag's motion is barely visible during small amplitude limit cycle flapping ($Ca=0.407$), and the wake is almost steady. Limit cycle flapping ($Ca=0.350$) and chaotic flapping ($Ca=0.164$) result in the ejection of vortex dipoles in the wake. Deflected mode flapping ($Ca=0.107$) generates a bluff body wake, where a typical \emph{von K\'arm\'an} vortex street is generated.\\

\begin{figure}
    \centering
    \def\svgwidth{\textwidth}
    \graphicspath{{Figures/}}
    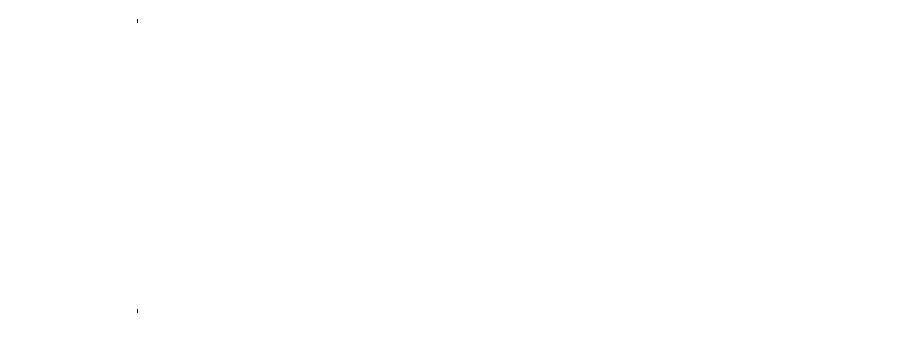
    \caption{Inverted flag of mass $M_\rho=0.5$ and $Re=200$ for various stiffness $Ca$; $(a)$ Bifurcation diagram of vertical tip displacement and $(b)$ peak frequency response of the different modes. Frequencies represented contain at least 20\% of the total energy spectrum. Points are coloured by Cauchy numbers.}
    \label{fig5}
\end{figure}

Figure~\ref{fig5} presents a bifurcation diagram of the flapping amplitude (taken at the zero-crossing of the velocities, corresponding to the maximum amplitude) and for the dominant flapping frequency for various Cauchy numbers. The bifurcation lines between the different flapping modes for this case are taken from \cite{Goza2018}. The peak frequency is the dominant flapping frequency from a Fourier transform of the vertical tip displacement, keeping only modes containing more than 20\% of the total energy. The transition between the different flapping regimes is exceptionally well captured, except for the $IV$-$V$ bifurcation that occurs earlier in our simulations. Similar observations can be made concerning the flapping frequencies. Specifically, for $Ca=0.350$, $Re=200$ and $M_\rho=0.5$, we find a maximum flapping amplitude and a dominant flapping frequency of $\delta_{\text{tip}}/L=0.80$ and $f=0.179$ Hz, respectively. These results match the ones of \cite{Goza2018} obtained using identical parameters, where they find a maximum flapping amplitude and a dominate flapping frequency of $\delta_{\text{tip}}/L=0.80$ and $f=0.181$ Hz, respectively. Limit cycle flapping results in a single dominant frequency, except close to the transition to the chaotic regime, where multiple dominant frequencies appear. The theoretical values of the dominant frequency of flapping obtained using small plate deflection theory and potential flow from \cite{Shen2016} is shown as a dotted line. The linear model overestimates the flapping frequency by a factor of 2 in the region of limit-cycle flapping as it relies on small amplitude deformation and a potential flow assumption. At lower Cauchy numbers, for deflected mode flapping, the dominant frequency agrees well with the theoretical bluff-body vortex shedding frequency of $f\sim0.2/\delta_{tip}$ (dashed line).

Additionally, we perform a simulation with a reduced mass ratio ($M_\rho=0.05$) to investigate the stability and efficiency of our coupled solver. As expected physically, we find little difference between the flapping regimes for these two distinct mass ratios. The partitioned solver requires more quasi-Newton iterations to reach the same residual (5.62 vs 3.56 per time step on average) but the method remains fully stable and accurate for this limiting mass ratio.

\subsubsection{Influence of Relaxation Methods}

Finally, we investigate the influence of the different iterative methods applied to the fluid-structure interaction problem. To allow constant relaxation to converge within a reasonable number of inner-iteration, the time step is reduced to $\Delta tU/\Delta x=0.1$, and the mass ratio is increased to $M_\rho=5$. The IQN-ILS method can use time steps twice as large and converge with fewer iterations than the constant relaxation method for this mass ratio, as demonstrated in the previous section. Results are presented in \tref{tab:relax} where we have simulated the inverted flag problem for 50 convective times.

\begin{table}
\centering
    \caption{Comparison of the different relaxation methods for the inverted flag problem with $Ca=0.350$, $M_\rho=5$ and $Re=200$ for a relative convergence measure of $\varepsilon_{rel}= 10^{-4}$. Aitken and IQN-ILS methods start with a constant under-relaxation of $\omega=0.05$.}
    \begin{tabular}{lllll}
    \toprule
   Method & avrg. iter. &  rel. iter. & rel. time (time)  & IQN/time\\
    \midrule
    % Explicit & & & & \\
    Constant-relaxation ($\omega=0.5$) & 10.3 & 100.0\% & 100.0\% (5180s) & -\\
    \midrule
    Aitken's relaxation & 4.2 & 40.4\% & 58.6\% (3010s) & -\\
    \midrule
    IQN-ILS (no initialisation) & 3.2 & 30.8\% &44.5\% (2303s) &0.4\%\\
    \midrule
    IQN-ILS  & 2.7 & 25.9\% & 18.5\% (958s) & 1.3\% \\
    \bottomrule
    \end{tabular}
    \label{tab:relax}
\end{table} 

Convergence within every time step is achieved for the quasi-Newton and the dynamic relaxation for this relatively simple problem. Constant relaxation does not converge in the first time step because of the large relaxation parameter required. Although iteration time is larger for the quasi-Newton scheme, the number of iterations per time step is drastically reduced, leading to an overall reduction in the computational time compared to standard relaxation. Additionally, proper initialisation of the coupling variable at the start of each time step improves convergence.

\section{Conclusion}

This article presented a novel Cartesian-grid fluid-structure interaction solver for thin, flexible structures. Based on our immersed boundary method, BDIM-$\sigma$ \cite{Lauber2022}, the solver specifically targets thin, dynamic structures. This immersed boundary code is coupled to a 3D shell solver \cite{chapelle2010finite} making the coupled solver very general. A strongly-coupled partitioned scheme is used to couple the two codes via the standard Dirichlet-Neumann approach. The interface quasi-Newton least-square method is used to efficiently update the interface which maintains accuracy and stability for strongly nonlinear problems while avoiding the expensive construction and solution of the full coupling-Jacobian.  

Through a series of examples for thin shell problems (up to $L/h=750$), we validate the current approach to predict the displacement and pressure distribution on a quasi-steady membrane wing, as well as the correct dynamical behaviour and bifurcation diagram of an inverted flag across a range of membrane stiffness. We show that the solver converges superlinearly ($p=1.6$) and remains stable for vanishing bending stiffness and vanishing structural mass. Finally, we show that the coupling method converges much faster than standard relaxation approaches, requiring only 18.5\% of the runtime.

\section*{Acknowledgments}
This work was financially supported by the UK Research and Innovation
Research Council Grant EP/L015382/1.

\printbibliography

\end{document}